\documentclass[aps,twocolumn,pra,raggedbottom,superscriptaddress,showpacs,tightenlines]{revtex4-1}
\usepackage{amsthm}
\usepackage{amsmath}
\usepackage{amssymb}
\usepackage{graphicx}
\usepackage{epstopdf}
\usepackage{url}
\usepackage{float}
\usepackage{bm}
\usepackage{ifthen}
\usepackage[usenames,dvipsnames]{color}
\usepackage{mathrsfs}
\usepackage{natbib}
\usepackage[colorlinks=true,citecolor=blue,urlcolor=blue]{hyperref}
\usepackage{paralist}
\usepackage{subfloat}
\usepackage{float}
\usepackage{braket}
\newcommand{\intd}[1]{\mathrm{d}#1}
\DeclareMathOperator{\Tr}{\text{Tr}}
\DeclareMathOperator{\Var}{\text{Var}}
\DeclareMathOperator{\Cov}{\text{Cov}}
\DeclareMathOperator{\LB}{\text{LB}}
\DeclareMathOperator{\UB}{\text{UB}}

\graphicspath{{Images/}} 
\begin{document}

\newcommand{\MITPhysics}{\affiliation{Department of Physics, Massachusetts Institute of Technology, Cambridge, Massachusetts 02139, USA}}
\newcommand{\RLE}{\affiliation{Research Laboratory of Electronics, Massachusetts Institute of Technology, Cambridge, Massachusetts 02139, USA}}

\pagestyle{empty}
\title{Practical high-dimensional quantum key distribution with decoy states}

\author{Darius Bunandar}
\email{dariusb@mit.edu}
\RLE \MITPhysics
\author{Zheshen Zhang}
\RLE
\author{Jeffrey H. Shapiro}
\RLE
\author{Dirk R. Englund}
\RLE
\date{\today}

\begin{abstract}
High-dimensional quantum key distribution (HD-QKD) allows two parties to generate multiple secure bits of information per detected photon. In this work, we show that decoy state protocols can be practically implemented for HD-QKD using only one or two decoy states. HD-QKD with two decoy states, under realistic experimental constraints, can generate multiple secure bits per coincidence at distances over 200 km and at rates similar to those achieved by a protocol with infinite decoy states. Furthermore, HD-QKD with only one decoy state is practical at short distances, where it is almost as secure as a protocol with two decoy states. HD-QKD with only one or two decoy states can therefore be implemented to optimize the rate of secure quantum communications.
\end{abstract}

\maketitle

\section{Introduction}
\label{sec:introduction}
High-dimensional quantum key distribution (HD-QKD), using qudits of dimensions $d>2$, enables its participants to optimize the secret-key capacity of a bosonic channel under technical constraints~\cite{2008.Zhang-Walmsley.QKD_CV}. When the secret-key generation rate is limited by the rate at which Alice generates photons or by the rate at which Bob can detect photons due to the detector dead time, the secret-key generation rate can be improved by high-dimensional photon encoding where each photon can encode as much as $\log_2 d > 1$ bits of information. Moreover, HD-QKD protocols may tolerate more noise than two-level, or qubit~\cite{BB84,1991.PRL.Ekert,1992.PRL.Bennett.BBM92}, QKD protocols~\cite{2002.PRL.Cerf-Gisin.d-levelQKD}.

Discrete HD-QKD protocols have been proven to be secure against coherent attacks, in which Eve is allowed to interact with all signals simultaneously~\cite{PRA.82.030301,PRA.83.039901}. Various photonic degrees of freedom have been investigated for HD-QKD, including position-momentum~\cite{2008.Zhang-Walmsley.QKD_CV}, time-energy~\cite{PRL.84.4737,PRL.93.010503,OptLett.31.2795,2007.PRL.Irfan-Howell.large_alphabet_QKD,2013.OpEx.Nunn.TFQKD}, transverse momentum~\cite{2013.SciRep.Etcheverry.HDQKD}, and orbital angular momentum~\cite{Nature.412.313,PRL.89.240401,PRL.92.167903,2013.PRA.Mafu.OAMQKD}. Among these, the time-energy basis is particularly attractive because time-energy correlations are compatible with wavelength-division multiplexing (WDM) systems and are robust in both free-space and fiber-based transmissions.

Recently, HD-QKD protocols employing time-energy entanglement have been proven to be secure against collective attacks, in which Eve's apparatus, which can include quantum memory, is restricted to interact with each signal separately~\cite{2013.PRA.Mower.DOQKD,2014.PRL.Zhang.FransonQKD}. (The bounds for unconditional security for both coherent and collective attacks turn out to be identical for most protocols~\cite{RevModPhys.81.1301}.) The proofs in~\cite{2013.PRA.Mower.DOQKD,2014.PRL.Zhang.FransonQKD} use the time-frequency covariance matrix (similar to the one used in continuous-variable QKD protocols~\cite{2003.Nature.Grosshans.CVQKD,PRL.97.190502,PRL.97.190503}) to derive a lower bound on the secure-key rate under collective attacks. The time-frequency covariance matrix can be measured using dispersive optics~\cite{1992.PRA.Franson.nonlocal_dispersion_cancellation,2013.PRA.Mower.DOQKD} or Franson interferometers~\cite{1989.PRL.Franson.Interferometer,2014.PRL.Zhang.FransonQKD}. The time-energy entanglement of photon pairs produced by spontaneous parametric down conversion (SPDC) has also been harnessed in several HD-QKD experiments~\cite{PRL.84.4737,PRA.73.031801,2007.PRL.Irfan-Howell.large_alphabet_QKD}.

All these experiments assume single-pair emissions from the SPDC source, whereas multi-pair emissions do occur. For a continuous-wave source, the signal and idler from each signal-idler mode pair are individually in identical thermal states with average photon numbers that are much smaller than 1. Therefore, when the HD-QKD frame time does not greatly exceed the source's correlation time, multi-pair emissions occurring during a particular frame will tend to be correlated in time, an effect known as photon bunching~\cite{1956.Nature.HBT.PhotonBunching}. In such cases, when any of these HD-QKD protocols is performed via a lossy channel, it is vulnerable to the photon number splitting (PNS) attack. On the other hand, when the frame time is much greater than the correlation time, the number of photon pairs emitted in a frame will be Poisson distributed, hence no photon bunching is then expected. Nevertheless, a PNS attack can provide Eve with some information about Alice and Bob's measurements when they reconcile their results via classical communication that Eve can monitor.

In the PNS attack, Eve measures the photon number of each transmission and selectively suppresses single photon signals~\cite{1995.PRA.Huttner.QKDwithCoherentStates,brassard00,2000.PRA.Lutkenhaus.PNSAttack,2002.NJP.Lutkenhaus.PNSAttack}. She then splits multiphoton signals---keeping one copy to herself and sending the other copy to Bob. Under the collective attack scheme, Eve stores her photons in a quantum memory and only measures them after Bob publishes his measurement bases over a public channel. She takes advantage of the timing correlations in the bunched photons to acquire information about Alice and Bob's key without being detected.

The decoy state protocol is designed to detect the PNS attack~\cite{2003.PRL.Hwang.DecoyQKD}. The central idea is to test the channel transmission properties by varying the source intensity. Decoy-state QKD has been discussed extensively in the context of Bennet-Brassard 1984 (BB84) protocol~\cite{2005.PRL.Wang.DecoyQKD,2005.PRL.Lo.DecoyQKD,2005.PRA.Ma.PracticalDecoy,2014.PRA.Lim.FinitePracticalDecoyQKD}. In addition, several experiments have demonstrated the generation of secure bits over 144 km in free space~\cite{PRL.98.010504} and over 107 km in optical fiber~\cite{PRL.98.010503}. Furthermore, it has been shown that decoy states can also be generated passively by using a beam splitter or by monitoring the idler of an SPDC source~\cite{2008.NJP.Ma.PassiveDecoySPDC, 2010.PRA.Curty.PassiveDecoyQKD, 2010.PRA.Curty.PassiveDecoyQKDSource, 2010.PRA.Xu.PassiveDecoyQKDwithPNR,2010.OpLett.Zhang.PassiveDecoyQKDExp, 2014.PRA.Krapick.BrightPassiveDecoyQKD, 2014.LPL.Sun.ExptPassiveDecoyState}. Recently, decoy-state analysis was extended to HD-QKD protocols~\cite{2014.PRL.Zhang.FransonQKD}, but for an infinite number of decoy states, which is practically impossible.

Here, we analyze the security of HD-QKD protocols employing a practical number of decoy states. Unlike the BB84 decoy-state QKD protocol, we make use of the decrease in measurement correlations instead of the quantum bit error rate (QBER) to estimate the amount of information gained by Eve. As a consequence, we find that the two-decoy-state protocol with one vacuum decoy state, which provides the best secure-key rate for BB84~\cite{2005.PRA.Ma.PracticalDecoy}, is not optimal for HD-QKD.

The analysis presented here answers a pressing question for experimental implementations of HD-QKD: how many decoy states are necessary for HD-QKD protocols to be robust against the PNS attack? We show by numerical evaluations, assuming realistic experimental parameters, that the security of a protocol with two decoy states approaches that of a protocol with an infinite number of decoy states. HD-QKD with only two decoy states can therefore be used to maximize the rate of high-speed secure quantum communications under experimental constraints.

We shall focus our discussion on a specific HD-QKD scheme: the dispersive optics QKD (DO-QKD) protocol~\cite{2013.PRA.Mower.DOQKD,2015.QIP.Lee.DOQKD,2014.PRA.Lee.DOQKD}, which employs group velocity dispersion to transform between mutually unbiased time and frequency bases. Although we restrict our analysis to DO-QKD, the same arguments are also applicable to other HD-QKD protocols employing time-energy entanglement.

This work is organized as follows:
Sec.~\ref{sec:DOQKD} briefly reviews the DO-QKD protocol.
Sec.~\ref{sec:generalDecoy} outlines the general decoy-state protocol. We discuss the relevant parameters that can be measured by Alice and Bob during quantum communication. In addition, we present a lower bound on the secure-key capacity when an infinite number of decoy states is available to Alice and Bob.
Sec.~\ref{sec:twoDecoy} derives a new lower bound on the secure-key capacity when only two decoy states are employed, and Sec.~\ref{sec:oneDecoy} considers the case of a single decoy state.
Sec.~\ref{sec:noDecoy} presents a lower bound on the secure-key capacity when no decoy state is employed.
The results of a numerical evaluation with realistic experimental constraints are presented in Sec.~\ref{sec:numerical}.
We defer the calculation of mutual information between Alice and Bob and the calculation of Eve's Holevo information to Appendices~\ref{sec:holevo} and~\ref{sec:shannon}.

\section{Dispersive optics quantum key distribution}
\label{sec:DOQKD}
\begin{figure}[h]
    \centering
    \includegraphics[width=\columnwidth]{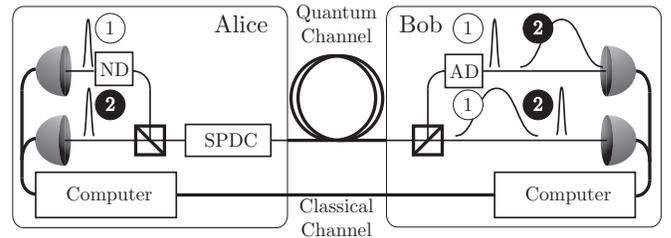}
    \caption{Schematic diagram of the DO-QKD setup. Alice and Bob randomly choose to measure in either the arrival-time basis or the frequency basis. In case 1, Alice measures in the frequency basis by applying a normal dispersion (ND). Bob's measurement is only anti-correlated to Alice's if he also measures in the frequency basis by applying an anomalous dispersion (AD). In case 2, Alice measures in the arrival-time basis, and Bob's measurement is only correlated to Alice's if he also measures in the arrival-time basis.}
\label{fig:DOQKDsetup}
\end{figure}

In the DO-QKD protocol, illustrated in Fig.~\ref{fig:DOQKDsetup}, Alice weakly pumps an SPDC source such that the time-energy entangled output state when only one pair is emitted can be approximated to have a Gaussian envelope \cite{PRA.73.031801}:
\begin{equation}
\psi(t_A, t_B) \propto e^{-(t_A-t_B)^2/4 \sigma_{\text{cor}}^2} e^{-(t_A + t_B)^2/16 \sigma_{\text{coh}}^2}.
\end{equation}
Here, $\sigma_{\text{coh}}$ is the coherence time of the pump field, and $\sigma_{\text{cor}}$ is the correlation time between the two photons generated by the SPDC source. $\sigma_{\text{coh}}$ typically can be longer than a microsecond for a diode laser, and $\sigma_{\text{cor}}$ is typically on the order of picoseconds for typical SPDC sources~\cite{2009.OpEx.Zhong-Battle.photon_pair_source}. The number of alphabet characters per photon pulse, $d = \sigma_{\text{coh}} / \sigma_{\text{cor}}$ (the Schmidt number), therefore can be large~\cite{2004.PRL.Law.Entanglement,2007.PRL.Irfan-Howell.large_alphabet_QKD}.

Alice and Bob randomly choose to measure their photons in the conjugate bases of photon arrival time and photon frequency; the two bases are measured using a fast single-photon detector or a dispersive optical element followed by photodetection, respectively. We assume that Alice and Bob have complete control of their own setups, precluding tampering by any third party such as Eve. In a single measurement frame, if both Alice and Bob measure their photons in the arrival-time basis, their timing measurements will be correlated. Similarly, if both parties measure in the frequency basis, their measurements will be anti-correlated. On the other hand, if one party measures in the frequency basis while the other measures in the arrival-time basis, the timing correlation between their photons is severely diminished.

After the measurement stage, Alice and Bob sift for frames in which both of them registered at least one detection event. For any frame with more than one coincidence, Alice and Bob replace their detection events with a random variable whose probability distribution matches that of photons originating from single-pair emissions. Finally, they apply error correction and privacy amplification to establish identical secret keys.

The DO-QKD protocol is not prone to the PNS attack when it is performed using an on-demand single photon source. When such a photon source is used, the bound on the secure-key capacity of a DO-QKD protocol, in terms of bits per  photon-pair coincidence (bpc), is~\cite{2005.ProcRoySoc.Devetak.Distillation,PRL.97.190503}
\begin{equation}
\label{eq:ideal}
\Delta I \geqslant \beta I(A;B) - \chi^{\UB}_{\xi_t,\xi_{\omega}}(A;E),
\end{equation}
where $\beta$ is the reconciliation efficiency and $I(A;B)$ is the mutual information between Alice and Bob. $\chi^{\UB}_{\xi_t,\xi_{\omega}}(A;E)$ is an upper bound on Eve's Holevo information under collective attacks, given the excess-noise factors $\xi_t$ and $\xi_{\omega}$ for the timing and the frequency correlations, respectively.

Eve's attack on Alice's transmission degrades the correlations of Alice and Bob's measurements in a manner parameterized by the excess-noise factors. Explicitly, $\Var[T'_A - T'_B] = (1+\xi_t) \Var[T_A - T_B]$ and $\Var[\Omega'_A + \Omega'_B] = (1+\xi_{\omega}) \Var[\Omega_A + \Omega_B]$, where $T_A$ ($T_B$) and $\Omega_A$ ($\Omega_B$) are the random variables associated with Alice's (Bob's) time and frequency measurements without Eve's presence. The corresponding primed variables are the random variables after Eve's intrusion. The sign difference is a consequence of Alice and Bob's timing measurements being directly correlated while their frequency measurements are anti-correlated. These excess-noise factors allow us to place an upper bound on Eve's Holevo information.

\section{General decoy state protocol}
\label{sec:generalDecoy}

\subsection{Postselection probability}
\label{sec:postselectionProbability}
We consider the practical case of interest for SPDC-based HD-QKD systems, i.e., we assume a continuous-wave source operating at low brightness (signal and idler beams have average photon numbers per mode much less than 1) with a frame time that greatly exceeds the correlation time. In this case, the photon-pair statistics are approximately Poissonian~\cite{2011.PRA.Ma.SinglePhotonMultiplex,2004.JMO.Riedmatten.SPDCStats}. Suppose that Alice's SPDC source emits an average of $\lambda$ pairs per measurement frame, the probability $\mathrm{Pr}_n$ of emitting $n$-photon pairs in a single measurement frame is then
\begin{equation}
    \mathrm{Pr}_n=\frac{\lambda^n}{n!}e^{-\lambda}.
\end{equation}
Furthermore, the postselection probability, which is the probability of Alice and Bob registering at least one detection (due to a photon or a dark count) in a single measurement frame, can be written as
\begin{equation}
\label{eq:postselectProb}
P_{\lambda} =  \sum^{\infty}_{n=0} \mathrm{Pr}_n C_n = \sum^{\infty}_{n=0} \frac{\lambda^n}{n!} e^{-\lambda} C_n,
\end{equation}
where $C_n$ is the conditional probability of measuring at least one detection given $n$-photon pairs are emitted. Explicitly, in Eve's absence we have
\begin{equation}
\label{eq:conditionalCoinProb}
\begin{aligned}
C_n = &\left[ 1 - (1-\eta_A)^n (1-p_d) \right] \\ \times
	  &\left[ 1 - (1-\eta_B \eta_P)^n (1-p_d) \right].
\end{aligned}
\end{equation}
Here, $\eta_A$ and $\eta_B$ are Alice and Bob's detector efficiencies, $\eta_P$ is the transmittance of the quantum channel linking Alice's source to Bob's terminal, and $p_d$ is the probability of one dark count in a single measurement frame. We are neglecting the possibility of multiple dark counts occurring in a frame because the product of the frame duration and the dark count rate for a typical superconducting nanowire single-photon detectors is much smaller than 1~\cite{2013.NPhoton.Marsili-Nam.93_perc_SNSPD}. Eve, in principle, has the freedom to affect the $C_n$ values. The goal of the decoy state protocol is to estimate the $C_n$ values from the postselection probabilities of different choices of $\lambda$.

\subsection{Excess noise}
\label{sec:excessNoise}
Alice and Bob cannot directly measure their timing and frequency correlations when there are multiphoton emissions and dark counts. They can only measure the averaged correlations:
\begin{equation}
\label{eq:totalVariances}
\begin{aligned}
\Var[T'_A - T'_B]_{\lambda} &= F_{\lambda} \Var[T'_A - T'_B] + (1-F_{\lambda}) \; \Delta \sigma^2_{t}, \\
\Var[\Omega_A' + \Omega_B']_{\lambda} &= F_{\lambda} \Var[\Omega_A' + \Omega_B'] + (1-F_{\lambda}) \; \Delta \sigma^2_{\omega},
\end{aligned}
\end{equation}
where $F_{\lambda} = \lambda e^{-\lambda} C_1/ P_{\lambda}$ is the fraction of postselected events that are due to single photon emissions, and $\Delta \sigma^2_t$ ($\Delta \sigma^2_{\omega}$) is the measured time (dispersed-time) correlations that are due to measurements of multiphoton emissions and dark counts.

It is convenient to divide~\eqref{eq:totalVariances} by $\Var[T_A - T_B]$ or $\Var[\Omega_A+\Omega_B]$ so that the excess-noise factors $\xi_t$ and $\xi_{\omega}$ are explicit:
\begin{equation}
\label{eq:noiseMultipliers}
\begin{aligned}
\Xi_{t,\nu} &= F_{\lambda} (1 + \xi_{t}) + (1-F_{\lambda}) \; \Delta \Xi_{t}, \\
\Xi_{\omega,\nu} &= F_{\lambda} (1 + \xi_{\omega}) + (1-F_{\lambda}) \; \Delta \Xi_{\omega}.
\end{aligned}
\end{equation}
The quantity $\Xi_{x,\lambda}$ (for $x=t$ or $\omega$) is the averaged excess-noise multiplier, which can be measured by Alice and Bob.

\subsection{Infinite number of decoy states}
\label{sec:infDecoy}
Now suppose that Alice and Bob choose a signal state with an expected photon-pair number $\mu$ and decoy states with expected photon-pair numbers $\nu_1, \nu_2, \dots, \nu_m$. Alice and Bob can then use the knowledge of the postselection probabilities $\mathcal{P} = \{P_{\mu}, P_{\nu_1}, \dots, P_{\nu_m}\}$ and the multipliers $\mathcal{K} = \{\Xi_{x,\mu},\Xi_{x,\nu_1},\dots,\Xi_{x,\nu_m}\}$ (for $x=t$ and $\omega$) to estimate the values of $C_n$ and $\xi_x$.

If we assume that $m \rightarrow \infty$, the key length is infinite, and the values of $C_n$ are linearly independent of each other, then Alice and Bob can determine the $C_n$ values to arbitrarily high confidence by measuring the set $\mathcal{P}$. Similarly, by measuring the set $\mathcal{K}$, they can determine $\xi_{x}$ to arbitrarily high confidence. Therefore, Alice and Bob can detect any attack by Eve that affects the values of $C_n$ and $\xi_x$~\cite{2005.PRL.Lo.DecoyQKD,2003.PRL.Hwang.DecoyQKD,2005.PRL.Wang.DecoyQKD}.

The bound on the secure-key capacity with $m \rightarrow \infty$ decoy states is~\cite{2014.PRL.Zhang.FransonQKD}
\begin{equation}
\label{eq:infiniteDecoy}
    \Delta I \geqslant \beta I(A;B) - \chi',
\end{equation}
where $\chi'$ is the amount of information assumed to be lost to Eve, defined as
\begin{equation}
\label{eq:effectiveHolevo}
    \chi' = (1-F_{\mu})\;n_R + F_{\mu}\;\chi^{\UB}_{\xi_t,\xi_{\omega}}(A;E).
\end{equation}
Here, $F_{\mu} = \mu e^{-\mu} C_1/ P_{\mu}$, and $n_R$ is the number of random bits shared between Alice and Bob when they use an error-correcting code employing an average of $n_{\text{ECC}}$ syndrome bits, which are revealed over the public channel. $\beta = (n_R - n_{\text{ECC}})/I(A;B)$ is the reconciliation efficiency. We have assumed that Alice and Bob can derive no security from multiphoton emissions. Note that when the photon source is an on-demand single photon source $(F_{\mu}=1)$, we recover~\eqref{eq:ideal}.

\section{Two decoy states}
\label{sec:twoDecoy}

When only a few decoy states are available, Alice and Bob cannot determine (to arbitrarily high confidence) the amount of information lost to Eve, $\chi'$. They can, however, provide a reasonable upper bound to $\chi'$ by using the following methods:
\begin{enumerate}
    \item Finding a lower bound on $F_{\mu}$, which estimates how close their photon source is to an ideal one, and
    \item Finding upper bounds on $\xi_t$ and $\xi_{\omega}$, which estimate Eve's Holevo information $\chi(A;E)$.
\end{enumerate}

We suppose that Alice and Bob choose fewer than three \emph{weak} decoy states with mean photon-pair numbers $\nu_1$ and $\nu_2$ that satisfy
\begin{equation}
\label{eq:assumptions}
\begin{aligned}
0 \leqslant \nu_2 < \nu_1, \\
\nu_1 + \nu_2 < \mu.
\end{aligned}
\end{equation}

\subsection{Lower bound on $F_{\mu}$}
The postselection probabilities of the two different states are given by
\begin{equation}
P_{\nu_1} = \sum^{\infty}_{n=0} C_n \frac{\nu_1^n}{n!} e^{-\nu_1},
\end{equation}
and
\begin{equation}
P_{\nu_2} = \sum^{\infty}_{n=0} C_n \frac{\nu_2^n}{n!} e^{-\nu_2}.
\end{equation}
As shown in~\cite{2005.PRA.Ma.PracticalDecoy}, we can find a lower bound on $C_1$ from the difference of the two postselection probabilities:
\begin{equation}
\label{eq:ineqC1}
\begin{aligned}
C_1 &\geqslant \frac{\mu}{\mu\nu_1 - \mu\nu_2 - \nu_1^2 + \nu_2^2} \bigg[ P_{\nu_1}e^{\nu_1} - P_{\nu_2} e^{\nu_2} \\
	&\qquad- \frac{\nu_1^2 - \nu_2^2}{\mu^2} \left( P_{\mu} e^{\mu} - C_0 \right) \bigg],
\end{aligned}
\end{equation}
where the inequality follows from the relation: $(\nu_1/\mu)^n - (\nu_2/\mu)^n \leqslant (\nu_1/\mu)^2 - (\nu_2/\mu)^2$ for $n\geqslant2$ which is true given~\eqref{eq:assumptions}. The above relation tells us that a lower bound on $C_0$ is needed to make use of~\eqref{eq:ineqC1}. One such bound is
\begin{equation}
\label{eq:C0LowerBound1}
C_0 \geqslant \frac{\nu_1 P_{\nu_2} e^{\nu_2} - \nu_2 P_{\nu_1} e^{\nu_1} }{\nu_1 - \nu_2},
\end{equation}
which follows from the assumption that $\nu_1 > \nu_2$.

Another lower bound on $C_0$ can be found using the assumption that Eve does not have access to both Alice and Bob's experimental setups. Since Alice owns the SPDC source, Eve cannot tamper with Alice's measurement of any output state generated by the source. When the source emits no photons, Alice's detector can only register a dark count, which occurs with probability $p_d$. Eve is allowed to do whatever she pleases with the vacuum state heading towards Bob, such as injecting photons into the channel. However, whatever she does cannot lower the probability of Bob registering a count to any value below $p_d$. Therefore, we conclude
\begin{equation}
\label{eq:C0LowerBound2}
C_0 \geqslant p_d^2.
\end{equation}

Combining~\eqref{eq:C0LowerBound1} and~\eqref{eq:C0LowerBound2} then gives
\begin{equation}
\label{eq:C0LowerBound}
C_0 \geqslant C_{0}^{\LB,\{\nu_1,\nu_2\}}  = \max \left\{ \frac{\nu_1 P_{\nu_2} e^{\nu_2} - \nu_2 P_{\nu_1} e^{\nu_1} }{\nu_1 - \nu_2}, p_d^2 \right\}.
\end{equation}

By using~\eqref{eq:ineqC1} and~\eqref{eq:C0LowerBound}, we find
\begin{equation}
\label{eq:FLowerBound}
\begin{aligned}
F_{\mu}   &= C_{1} \frac{\mu e^{-\mu}}{P_{\mu}} \\
	&\geqslant \frac{\mu^2}{\mu \nu_1 - \mu \nu_2 - \nu_1^2 + \nu_2^2} \bigg[\frac{P_{\nu_1}}{P_{\mu}} e^{\nu_1-\mu} - \frac{P_{\nu_2}}{P_{\mu}} e^{\nu_2-\mu} \\
	&\qquad -\frac{\nu_1^2 - \nu_2^2}{\mu^2} \left( 1 - \frac{C_0^{\LB,\{\nu_1,\nu_2\}}e^{-\mu}}{P_{\mu}} \right) \bigg].
\end{aligned}
\end{equation}

Another way of obtaining a lower bound on $F_{\mu}$ is immediately evident from the postselection probability of a single decoy state. Let $\lambda = \nu_1$ or $\nu_2$ if $\nu_2 \neq 0$, and $\lambda = \nu_1$ if $\nu_2 = 0$. It then follows that
\begin{equation}
\begin{aligned}
P_{\lambda} e^{\lambda}
        &= C_0 + C_1 \lambda + \sum^{\infty}_{n=2} \frac{\lambda^n}{n!} C_n \\
        &< C_0 + C_1 \lambda + \frac{\lambda^2}{\mu^2} \sum^{\infty}_{n=2} \frac{\mu^n}{n!} C_n \\
        &= C_0 + C_1 \lambda + \frac{\lambda^2}{\mu^2} \left( P_{\mu}e^{\mu} - C_0 - C_1 \mu \right),
\end{aligned}
\end{equation}
because $\lambda/\mu \leqslant 1$. Solving for $C_1$ we obtain
\begin{equation}
C_1 > \frac{\mu}{\mu \lambda - \lambda^2} \left[ P_{\lambda} e^{\lambda} -  \frac{\lambda^2}{\mu^2} P_{\mu} e^{\mu}- \frac{\mu^2-\lambda^2}{\mu^2} C_0 \right],
\end{equation}
which is similar to what is found in Ref.~\cite{2005.PRA.Ma.PracticalDecoy} using another method.

Now, we need to upper bound $C_0$ to find the lower bound of $C_1$. We again assume that Eve cannot intrude into Alice and Bob's experimental setups. This implies that, when Alice's source emits no photons, Alice and Bob's conditional coincidence probability $C_0$ cannot exceed the dark count probability of Alice's detectors:
\begin{equation}
C_0 \leqslant C_0^{\UB,\{\nu_1,\nu_2\}} = p_d.
\end{equation}
Therefore,
\begin{equation}
\begin{aligned}
\label{eq:FLowerBound2}
F_{\mu} &= C_1 \frac{\mu e^{-\mu}}{P_{\mu}} \\
    &> \frac{\mu^2}{\mu \lambda - \lambda^2} \left[ \frac{P_{\lambda}}{P_{\mu}} e^{\lambda-\mu} -  \frac{\lambda^2}{\mu^2} - \frac{\mu^2-\lambda^2}{\mu^2}\frac{C_0^{\UB,\{\nu_1,\nu_2\}} e^{-\mu}}{P_{\mu}} \right],
\end{aligned}
\end{equation}
where  $\lambda = \nu_1$ or $\nu_2$ if $\nu_2 \neq 0$, and $\lambda = \nu_1$ if $\nu_2 = 0$.

Combining~\eqref{eq:FLowerBound} and~\eqref{eq:FLowerBound2}, we get
\begin{equation}
\begin{aligned}
\label{eq:FLowerBoundTwoDecoy}
F_{\mu} &\geqslant F_{\mu}^{\LB,\{\nu_1,\nu_2\}} \\
    &= \max \left\{ \frac{\mu^2}{\mu \nu_1 - \mu \nu_2 - \nu_1^2 + \nu_2^2} \bigg[\frac{P_{\nu_1}}{P_{\mu}} e^{\nu_1-\mu} - \frac{P_{\nu_2}}{P_{\mu}} e^{\nu_2-\mu}\right. \\
    &\qquad -\frac{\nu_1^2 - \nu_2^2}{\mu^2} \left( 1 - \frac{C_0^{\LB,\{\nu_1,\nu_2\}}e^{-\mu}}{P_{\mu}} \right) \bigg], \\
    &\left. \frac{\mu^2}{\mu \lambda - \lambda^2} \left[ \frac{P_{\lambda}}{P_{\mu}} e^{\lambda-\mu} -  \frac{\lambda^2}{\mu^2} - \frac{\mu^2-\lambda^2}{\mu^2}\frac{C_0^{\UB,\{\nu_1,\nu_2\}} e^{-\mu}}{P_{\mu}} \right]\right\},
\end{aligned}
\end{equation}
where  $\lambda = \nu_1$ or $\nu_2$ if $\nu_2 \neq 0$, and $\lambda = \nu_1$ if $\nu_2 = 0$.

\subsection{Upper bounds on $\xi_{t}$ and $\xi_{\omega}$}

Let $(\lambda_1,\lambda_2) \in \mathcal{L} = \{(\mu,\nu_1),(\mu,\nu_2),(\nu_1,\nu_2)\}$. Each member of $\mathcal{L}$ is an ordered pair of two mean photon-pair numbers. The averaged excess-noise multipliers for the ordered pair $(\lambda_1,\lambda_2)$ are
\begin{equation}
\begin{aligned}
\Xi_{x,\lambda_1} = F_{\lambda_1} (1+\xi_x) + \Delta \Xi_{x} \left( 1-F_{\lambda_1} \right), \\
\Xi_{x,\lambda_2} = F_{\lambda_2} (1+\xi_x) + \Delta \Xi_{x} \left( 1-F_{\lambda_2} \right).
\end{aligned}
\end{equation}
Multiplying the above two equations by $P_{\lambda_1} e^{\lambda_1}$ and $P_{\lambda_2} e^{\lambda_2}$ respectively, we obtain
\begin{equation}
\begin{aligned}
\Xi_{x,\lambda_1} P_{\lambda_1} e^{\lambda_1} = \lambda_1 C_1 (1+\xi_x) + \Delta \Xi_{x} \left( P_{\lambda_1} e^{\lambda_1} - \lambda_1 C_1 \right), \\
\Xi_{x,\lambda_2} P_{\lambda_2} e^{\lambda_2} = \lambda_2 C_1 (1+\xi_x) + \Delta \Xi_{x} \left( P_{\lambda_2} e^{\lambda_2} - \lambda_2 C_1 \right).
\end{aligned}
\end{equation}
To find upper bounds on $\xi_t$ and $\xi_{\omega}$, we take the difference between these two equations,
\begin{equation}
\begin{aligned}
&\Xi_{x,\lambda_1} P_{\lambda_1} e^{\lambda_1} - \Xi_{x,\lambda_2} P_{\lambda_2} e^{\lambda_2} \\
	&\qquad= (\lambda_1-\lambda_2) C_1 (1+\xi_x) \\
	&\qquad\quad + \Delta \Xi_{x} \left( P_{\lambda_1} e^{\lambda_1} - P_{\lambda_2} e^{\lambda_2} -(\lambda_1-\lambda_2) C_1 \right) \\
	&\qquad\geqslant (\lambda_1-\lambda_2) C_1 (1+\xi_x),
\end{aligned}
\end{equation}
where the inequality comes from
\begin{equation}
\begin{aligned}
P_{\lambda_1} e^{\lambda_1} - P_{\lambda_2} e^{\lambda_2} &= \sum^{\infty}_{n=0} \frac{\lambda_1^n - \lambda_2^n}{n!} C_n \\
	&\geqslant (\lambda_1 - \lambda_2) C_1,
\end{aligned}
\end{equation}
since $\lambda_1 > \lambda_2$ for any ordered pair $(\lambda_1,\lambda_2) \in \mathcal{L}$.
Thus,
\begin{equation}
\label{eq:xiUpperBound1}
\begin{aligned}
(1+\xi_x) &\leqslant \frac{1}{(\lambda_1 - \lambda_2) C_{1}} \left(\Xi_{x,\lambda_1} P_{\lambda_1} e^{\lambda_1} - \Xi_{x,\lambda_2} P_{\lambda_2} e^{\lambda_2}\right) \\
	&\leqslant 	\frac{\mu e^{-\mu}}{(\lambda_1-\lambda_2) F_{\mu}^{\LB,\{\nu_1,\nu_2\}}} \\
    &\qquad \times \left(\Xi_{x,\lambda_1} \frac{P_{\lambda_1}}{P_{\mu}} e^{\lambda_1} - \Xi_{x,\lambda_2} \frac{P_{\lambda_2}}{P_{\mu}} e^{\lambda_2}\right),
\end{aligned}
\end{equation}
for $x = t$ and $\omega$.

Another way to place upper bounds on $\xi_t$ and $\xi_{\omega}$ is immediately evident from~\eqref{eq:noiseMultipliers}:
\begin{equation}
\begin{aligned}
\Xi_{x,\lambda} &=  F_{\lambda} \; (1 + \xi_{x}) + \left(1- F_{\lambda}\right) \; \Delta \Xi_{x} \\
		&\geqslant F_{\lambda} (1 + \xi_{x}) \\
		&= \frac{\lambda P_{\mu}}{\mu P_{\lambda}} e^{\mu-\lambda} F_{\mu} \; (1 + \xi_{x}) \\
		&\geqslant \frac{\lambda P_{\mu}}{\mu P_{\lambda}} e^{\mu-\lambda} F_{\mu}^{\LB,\{\nu_1,\nu_2\}} \; (1 + \xi_{x}),
\end{aligned}
\end{equation}
for $\lambda \in \{\mu$, $\nu_1$, $\nu_2\}$. The inequality above implies that
\begin{equation}
\label{eq:xiUpperBound2}
(1+\xi_{x}) \leqslant  \min_{\lambda \in\{\mu,\nu_1,\nu_2\}}
	\left\{ e^{\lambda-\mu} \frac{\mu P_{\lambda}}{\lambda P_{\mu}} \frac{\Xi_{x,\lambda}}{F_{\mu}^{\LB,\{\nu_1,\nu_2\}}} 	\right\}.
\end{equation}
Combining~\eqref{eq:xiUpperBound1} and~\eqref{eq:xiUpperBound2} gives us
\begin{equation}
\label{eq:xiUpperBound}
\begin{aligned}
\xi_x &\leqslant \xi^{\UB,\{\nu_1,\nu_2\}}_{x} \\
	&= \min \bigg\{ \min_{(\lambda_1,\lambda_2) \in \mathcal{L}} \left\{\frac{\mu e^{-\mu}}{(\lambda_1-\lambda_2) F_{\mu}^{\LB,\{\nu_1,\nu_2\}}}\right.  \\
	& \qquad \left.\times \left(\Xi_{x,\lambda_1} \frac{P_{\lambda_1}}{P_{\mu}} e^{\lambda_1} - \Xi_{x,\lambda_2} \frac{P_{\lambda_2}}{P_{\mu}} e^{\lambda_2} \right)\right\}, \\
	& \qquad \min_{\lambda \in\{\mu,\nu_1,\nu_2\}}
	\left\{ e^{\lambda-\mu} \frac{\mu P_{\lambda}}{\lambda P_{\mu}} \frac{\Xi_{x,\lambda}}{F_{\mu}^{\LB,\{\nu_1,\nu_2\}}} \right\} \bigg\} -1.
\end{aligned}
\end{equation}

Using~\eqref{eq:FLowerBound} and~\eqref{eq:xiUpperBound}, we obtain a bound on the secure-key capacity of HD-QKD using only two decoy states:
\begin{equation}
\label{eq:twoDecoy}
\begin{aligned}
\Delta I \geqslant &\beta I(A;B)_{\mu} - (1-F_{\mu}^{\LB,\{\nu_1,\nu_2\}}) n_R \\
	&- F_{\mu}^{\LB,\{\nu_1,\nu_2\}} \chi^{\UB}_{\xi^{\UB,\{\nu_1,\nu_2\}}_t,\xi^{\UB,\{\nu_1,\nu_2\}}_{\omega}}(A;E),
\end{aligned}
\end{equation}
where the subscript $\mu$ on $I(A;B)$ indicates that Alice and Bob's mutual information is calculated using the signal state.

\section{One decoy state}
\label{sec:oneDecoy}

When Alice only uses one decoy state, whose mean photon-pair number $\nu$ is smaller than that of the signal state $\mu$, we can find a lower bound on $F_{\mu}$ by using~\eqref{eq:FLowerBound2} with $\lambda = \nu$. The argument used to upper bound $C_0$ still applies because it only depends on the assumption that Eve cannot intrude into Alice and Bob's experimental setups. Therefore,
\begin{equation}
\label{eq:FLowerBoundOneDecoy}
\begin{aligned}
F_{\mu} &\geqslant F_{\mu}^{\LB,\{\nu\}} \\
    &= \frac{\mu^2}{\mu \nu - \nu^2} \left[ \frac{P_{\nu}}{P_{\mu}} e^{\nu-\mu} -  \frac{\nu^2}{\mu^2} - \frac{\mu^2-\nu^2}{\mu^2}\frac{C_0^{\UB,\{\nu\}} e^{-\mu}}{P_{\mu}} \right],
\end{aligned}
\end{equation}
with $C_0 \leqslant C_0^{\UB,\{\nu\}} = p_d$.

Similarly, upper bounds on $\xi_t$ and $\xi_{\omega}$ can be found by using~\eqref{eq:xiUpperBound} with $(\lambda_1,\lambda_2) = (\mu,\nu)$ and $\lambda \in \{\mu,\nu\}$:
\begin{equation}
\label{eq:xiUpperBoundOneDecoy}
\begin{aligned}
\xi_{x} &\leqslant \xi^{\UB,\{\nu\}}_{x} \\
&=  \min \left\{ \frac{\mu}{(\mu-\nu) F_{\mu}^{\LB,\{\nu\}}}\left( \Xi_{x,\mu} - \frac{P_{\nu}}{P_{\mu}} \Xi_{x,\nu} e^{\nu-\mu} \right) \right., \\
& \qquad \left.\min_{\lambda \in\{\mu,\nu\}}
    \left\{ e^{\lambda-\mu} \frac{\mu P_{\lambda}}{\lambda P_{\mu}} \frac{\Xi_{x,\lambda}}{F_{\mu}^{\LB,\{\nu\}}} \right\} \right\} -1,
\end{aligned}
\end{equation}
for $x = t$ and $\omega$.

\section{No decoy states}
\label{sec:noDecoy}
When decoy states are not employed, Alice and Bob must use a fraction of their signal frames to estimate the transmission parameters. To find a lower bound on $F_{\mu}$, consider
\begin{equation}
\begin{aligned}
P_{\mu} e^{\mu} &= C_0 + C_1 \mu + \sum^{\infty}_{n=2} \frac{\mu^n}{n!} C_n \\
			&\leqslant C_0^{\UB,\emptyset} + C_1 \mu + \sum^{\infty}_{n=2} \frac{\mu^n}{n!}C_n^{\UB,\emptyset},
\end{aligned}
\end{equation}
where
\begin{equation}
C_n \leqslant C_n^{\UB,\emptyset} = 1 - (1-\eta_A)^n(1-p_d),
\end{equation}
is a consequence of Eve's inability to affect Alice's detection probability.

Using the relations above, we have
\begin{equation}
C_1 \geqslant C^{\LB,\emptyset}_1 = \frac{1}{\mu} \left[P_{\mu} e^{\mu} -  C_0^{\UB,\emptyset} - \sum^{\infty}_{n=2} \frac{\mu^n}{n!} C_n^{\UB,\emptyset} \right],
\end{equation}
and hence
\begin{equation}
\begin{aligned}
\label{eq:FLowerBoundNoDecoy}
F_{\mu}  &\geqslant F_{\mu}^{\LB,\emptyset} \\
	&= C^{\LB,\emptyset}_1 \frac{\mu e^{-\mu}}{P_{\mu}} \\
	&= 1 - \frac{ C_0^{\UB,\emptyset} e^{-\mu}}{P_{\mu}} - \sum^{\infty}_{n=2} \frac{\mu^n}{n!} \frac{C_n^{\UB,\emptyset} e^{-\mu}}{P_{\mu}}.
\end{aligned}
\end{equation}
Because $\Xi_{x,\mu}$ is the only available excess-noise multiplier, the upper bounds on $\xi_t$ and $\xi_{\omega}$ are found by using~\eqref{eq:xiUpperBound2} with $\lambda = \mu$:
\begin{equation}
\label{eq:xiUpperBoundNoDecoy}
\begin{aligned}
\xi_{x} &\leqslant \xi^{\UB,\emptyset}_{x} =  \frac{\Xi_{x,\mu}}{F_{\mu}^{\LB,\emptyset}} - 1,
\end{aligned}
\end{equation}
for $x = t$ and $\omega$.

\section{Numerical results and discussion}
\label{sec:numerical}

\begin{figure*}[t]
        \centering
        \includegraphics[width=1.5\columnwidth]{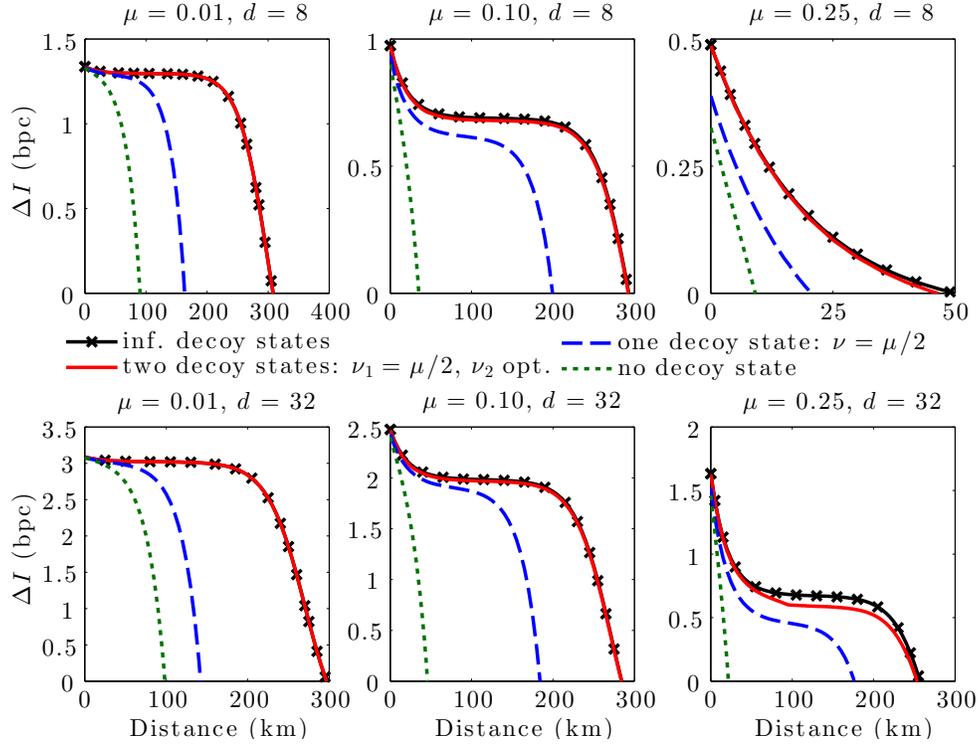}
    \caption{Lower bounds on the secure-key capacity (in bits per coincidence) of decoy-state HD-QKD as a function of transmission distance. Top panels show the case $d=8$ and bottom panels show the case $d=32$. Solid lines with crosses (black) correspond to HD-QKD with infinite decoy states; solid lines (red) correspond to HD-QKD with two weak decoy states of $\nu_1 = \mu/2$ and an optimized $\nu_2$; dashed lines (blue) correspond to HD-QKD with only one decoy state of $\nu = \mu/2$; and dotted lines (green) show the performance of HD-QKD without decoy states. For $\mu = 0.01$ and $0.10$ ($d = 8$ and $32$), lines for the infinite-decoy-state and the two-decoy-state protocols are indistinguishable at the plots' scales.}
\label{fig:decoyDOQKD}
\end{figure*}

\begin{figure}[h]
        \centering
        \includegraphics[width=1\columnwidth]{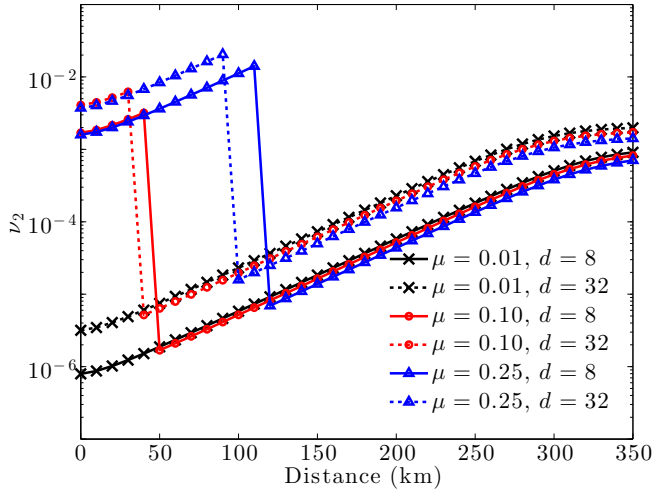}
    \caption{Optimal values of $\nu_2$ at different transmission distances for two-decoy-state protocols with $\mu=\{0.01,0.10,0.25\}$ and $\nu_1 = \mu/2$.}
\label{fig:optimizedNu2}
\end{figure}

Figure~\ref{fig:decoyDOQKD} plots the secure-key capacity of decoy-state HD-QKD with an SPDC source of mean photon-pair numbers per frame $\mu = 0.01,0.10$, and $0.25$. The top panels show the case in which the Schmidt number $d=8$ while the bottom panels show the case in which $d=32$. Three different decoy state protocols are plotted in each panel: the one-decoy-state protocol, the two-decoy-state protocol, and the infinite-decoy-state protocol. For comparison, we also plot the security of HD-QKD protocol without decoy states.

In particular, we consider the case $\nu = \mu/2$ for the one-decoy-state protocol. For the two-decoy-state protocol, we similarly assume $\nu_1 = \mu/2$, but we optimize $\nu_2$ such that, for any particular transmission distance, the lower bound on the secure-key capacity $\Delta I$ is maximized. Figure~\ref{fig:optimizedNu2} plots the optimal values of $\nu_2$ as a function of transmission distance at 10 km increments.

For the cases of $\mu = 0.10$ and $0.25$, ~\eqref{eq:FLowerBound2} gives a better lower bound on $F_{\mu}$ at short distances. The sharp drop in the optimal values of $\nu_2$ (at $\sim$50 km for $\mu=0.10$ and at $\sim$100 km for $\mu=0.25$) indicates where~\eqref{eq:FLowerBound} starts to provide a better lower bound on $F_{\mu}$ than~\eqref{eq:FLowerBound2}. On the other hand, for the cases of $\mu = 0.01$,~\eqref{eq:FLowerBound} provides a better lower bound on $F_{\mu}$ at all distances. Moreover, the optimal values of $\nu_2$ are small compared to $\mu$---but non-zero. This result is in contrast to the two-decoy-state BB84 protocol whose lower bound on secure-key capacity is always maximized when $\nu_2 \rightarrow 0$~\cite{2005.PRA.Ma.PracticalDecoy}.

We take $\sigma_{\text{cor}}$ = 30 ps for both $d$ values, and $\sigma_{\text{coh}} = d \sigma_{\text{cor}}$. The frame duration $T_f$ is chosen to be $T_f = 2 \sqrt{2 \ln{2}} \; \sigma_{\text{coh}}$. Experimentally, when a larger $d$ is wanted, it is easier to increase the coherence time $\sigma_{\text{coh}}$ than to decrease the correlation time $\sigma_{\text{cor}}$. This is because the $\sigma_{\text{coh}}$ can be increased by modulating the pulse duration of the laser pump field. On the other hand, $\sigma_{\text{cor}}$ is determined by the phase-matching bandwidth of the SPDC source and is characteristic to the parametric down-conversion process.

We assume the following experimental parameters: propagation loss $\alpha$ = 0.2 dB/km; detector timing jitter $\sigma_J$ = 20 ps; dark count rate $r_D = 1000$ s$^{-1}$; reconciliation efficiency $\beta$ = 0.9; $n_R = \log_2{d}$. The transmittance $\eta_P = 10^{-\alpha L/10}$, where $L$ is the length of the quantum channel in km. We also assume that Alice and Bob have the same detector efficiencies: $\eta_A = \eta_B = 0.93$~\cite{2013.NPhoton.Marsili-Nam.93_perc_SNSPD}.

For simplicity, we assume equal excess-noise factors for both the arrival-time and the frequency measurements, $\xi_t = \xi_{\omega} = \xi$. The change in correlation time due to Eve's interaction is assumed to be $\sigma_{\Delta} = \left( \sqrt{1+\xi} -1\right) \times \sigma_{\text{cor}} = 10$ ps. When Alice and Bob do not use an infinite number of decoy states, they can only measure $\Xi_{\mu}$. For the calculations, we assume that $\Delta \Xi = 1 + \xi$. Details on calculating Alice and Bob's mutual information, as well as Eve's Holevo information, are outlined in Appendices~\ref{sec:holevo} and~\ref{sec:shannon}.

Using decoy states improves the security of the HD-QKD protocol. For example, while the case of $\mu = 0.25$ and $d=32$ is insecure beyond 25 km without decoy states, the one-decoy-state protocol is able to generate 0.45 secure bpc at a distance of 100 km. Furthermore, when two weak decoy states are used, the protocol can generate more than 0.52 secure bpc up to a distance of 200 km.

Even though the probability of multiphoton emissions is low for $\mu = 0.01$, we only obtain secure bits up to a distance of $\sim$100 km without decoy states. However, the presence of one decoy state allows us to obtain 1.22 secure bpc for $d=8$ and 2.57 secure bpc for $d=32$ at the 100 km distance. The two-decoy-state protocols generate more than 1.26 secure bpc for $d=8$ and more than 2.83 secure bpc for $d=32$ up to a distance of 200 km.

In Fig.~\ref{fig:decoyDOQKD}, we also see that protocols with two decoy states perform almost as well as protocols with infinite decoy states. Intuitively, a protocol with an infinite number of decoy states should perform the best because an infinite number of decoy states allows us to estimate the values of all $C_n$ precisely. Nevertheless, the two-decoy-state protocols asymptotes to the infinite-decoy-state protocols, performing only slightly worse in the generation of secure-bit capacities at similar transmission distances. When two decoy states are employed, Alice and Bob can find useful lower bounds on $C_1$ and $C_0$ (and hence $F_{\mu}$). High-dimensional QKD protocols with two decoy states therefore appear practical as they offer multiple secure bits per coincidence at distances and at rates similar to those achieved by a protocol with infinite decoy states.

The two-decoy-state protocol can reach a longer secure distance than the one-decoy-state protocol. To see why, consider~\eqref{eq:postselectProb} and~\eqref{eq:conditionalCoinProb}. Notice that at short distances, where the transmittance $\eta_P \sim 1$, the postselection probability is dominated by $C_n$ with small values of $n$. However, at large distances, where the transmittance $\eta_P \ll 1$, the postselection probability is dominated by $C_n$ with large values of $n$. Therefore, referring to~\eqref{eq:FLowerBoundOneDecoy}, the lower bound on $F_{\mu}$ in the one-decoy-state protocol, calculated by taking the difference between $P_{\nu} e^{\nu}$ and $C_0$, decreases quickly as the channel transmittance $\eta_P$ decreases. On the other hand, the lower bound of $F_{\mu}$ in~\eqref{eq:FLowerBound} for the two-decoy-state protocol is calculated by taking the difference between $P_{\nu_1} e^{\nu_1}$ and $P_{\nu_2} e^{\nu_2}$, which are of comparable values at both short and long distances. The one-decoy-state protocol is nevertheless easy to implement. Moreover, the one-decoy-state protocol offers boosts to the lower bound on secure-key capacity, increasing the secure distance and the generation rate, of a protocol without decoy states.

It is also interesting that, independent of the number of decoy states employed, the photon efficiency of HD-QKD (in bpc) decreases rapidly with increasing $\mu$. The case of $\mu = 0.25$ and $d=8$ is insecure at only 50 km, even when infinite decoy states are used. This implies that the $\mu$ value employed in HD-QKD should be chosen to ensure that the probability of multiphoton emissions is low.

\section{Conclusion}
\label{sec:conclusion}

We have analyzed the practicality of HD-QKD protocols with decoy states. In particular, we considered the case of HD-QKD with two decoy states and with one decoy state. For completeness, we have also studied how the HD-QKD would perform without decoy states.

Through simple numerical examples, we have shown that HD-QKD with two decoy states is practical: it can achieve multiple secure bits per coincidence at distances over 200 km and at rates similar to those achieved by a protocol with infinite decoy states. The HD-QKD protocol with only one decoy state is also practical at short distances, in which case it is almost as secure as the two-decoy-state protocol at short distances.

While we have only considered the DO-QKD protocol, the arguments presented in this work can be generalized to other HD-QKD protocols~\cite{2014.PRL.Zhang.FransonQKD,2013.OpEx.Nunn.TFQKD}. Decoy-state HD-QKD protocols that are robust against collective PNS attacks can therefore be used to maximize the rate of high-speed secure quantum communications.

\section{Acknowledgments}
\label{sec:acknowledgments}

The authors would like to thank Jacob Mower, Catherine Lee, and Gregory Steinbrecher for their helpful discussions. This work was supported by the DARPA Quiness Program through U.S. Army Research Office Grant No. W31P4Q-12-1-0019. DB acknowledges the support of the Bruno Rossi Graduate Fellowship in Physics at MIT.

\appendix
\section{Eve's Holevo information}
\label{sec:holevo}
The output state from an SPDC source in the low-flux limit is Gaussian, and Gaussian attacks are optimal for a given covariance matrix~\cite{PRL.97.190503,PRL.97.190502}. Alice and Bob's time-frequency covariance matrix is therefore crucial in estimating Eve's Holevo information~\cite{RevModPhys.84.621}. Before any interaction with Eve, it is
\begin{equation}
\Gamma =
\begin{pmatrix}
\gamma_{AA}	&	\gamma_{AB}	\\
\gamma_{BA}	&	\gamma_{BB}
\end{pmatrix},
\end{equation}
where the submatrices $\gamma_{JK}$ for $J,K=A,B$ are given by
\begin{equation}
\begin{aligned}
\gamma_{AA} &=
\begin{pmatrix}
\frac{u+v}{16}		&	-\frac{u+v}{8k}	\\
-\frac{u+v}{8k}		&	\frac{(u+v)(4k^2+uv)}{4k^2uv}
\end{pmatrix}, \\
\gamma_{AB} &= \gamma_{BA}^T =
\begin{pmatrix}
\frac{u-v}{16}		&	\frac{u-v}{8k}	\\
-\frac{u-v}{8k}		&	-\frac{(u-v)(4k^2+uv)}{4k^2uv}
\end{pmatrix}, \\
\gamma_{BB} &=
\begin{pmatrix}
\frac{u+v}{16}		&	\frac{u+v}{8k}	\\
\frac{u+v}{8k}		&	\frac{(u+v)(4k^2+uv)}{4k^2uv}
\end{pmatrix}, \\
\end{aligned}
\end{equation}
with $u = 16 \sigma^2_{\text{coh}}$ and $v = 4 \sigma^2_{\text{cor}}$~\cite{2013.PRA.Mower.DOQKD}. Note that every entry in the covariance matrix is measured in units of time. After Eve's interaction, the new covariance matrix is
\begin{equation}
\Gamma' =
\begin{pmatrix}
\gamma'_{AA}	&	\gamma'_{AB}	\\
\gamma'_{BA}	&	\gamma'_{BB}
\end{pmatrix},
\end{equation}
where the new submatrices are
\begin{equation}
\begin{aligned}
\gamma_{AA}' &= \gamma_{AA}, \\
\gamma_{AB}' &= (\gamma_{BA}')^T =
\begin{pmatrix}
1-\eta_t	&	0				\\
0		&	1-\eta_{\omega}
\end{pmatrix} \gamma_{AB}, \\
\gamma_{BB}' &=
\begin{pmatrix}
1-\epsilon_t	&	0				\\
0		&	1-\epsilon_{\omega}
\end{pmatrix} \gamma_{BB}. \\
\end{aligned}
\end{equation}
Here, $\eta_t$ and $\eta_{\omega}$ represent the decrease in correlations, while $\epsilon_t$ and $\epsilon_{\omega}$ represent the excess noise---all due to Eve's interactions.

Once Alice and Bob have estimated the covariance matrix $\Gamma'$, we can then assume that Alice, Bob, and Eve share a pure Gaussian state $\rho_{ABE}$ in evaluating Eve's Holevo information. If Alice and Bob only generate secure bits from their arrival-time measurements, Eve's Holevo information can then be calculated from
\begin{equation}
\chi_{\xi_t,\xi_{\omega}} (A;E) = S(\rho_{AB}) - S(\rho_{\left. B \right| T_A}),
\end{equation}
where $S(\rho) = -\Tr[\rho \log_2 \rho]$ is the von Neumann entropy of the quantum state $\rho$. $S(\rho_{AB})$ can then be evaluated from $S(\rho_{AB}) = f(d_{+}) + f(d_{-})$ where
\begin{equation}
f(x) = \left(x + \frac{1}{2} \right) \log_2 \left(x + \frac{1}{2}\right) - \left(x - \frac{1}{2} \right) \log_2 \left(x - \frac{1}{2}\right),
\end{equation}
and
\begin{equation}
\begin{aligned}
d_{\pm} &= \frac{1}{\sqrt{2}} \sqrt{I_1 \pm \sqrt{I_1^2 - 4I_2}}, \\
I_1 &= \det[\gamma'_{AA}] + \det[\gamma'_{BB}] + 2 \det[\gamma'_{AB}], \\
I_2 &= \det{\Gamma'}.
\end{aligned}
\end{equation}
Furthermore, $S(\rho_{\left. B \right| T_A})$ can be computed from
\begin{equation}
\begin{aligned}
S(\rho_{\left. B \right| T_A}) &= f \left(\sqrt{\det[\gamma'_{\left. B \right| T_A}]}\right), \\
\end{aligned}
\end{equation}
where
\begin{equation}
\begin{aligned}
\gamma'_{\left. B \right| T_A} &= \gamma'_{BB} - \gamma'_{BA} \left(X_t \gamma'_{AA} X_t \right)^{-1} \gamma'_{AB}, \\
\end{aligned}
\end{equation}
Here,
$X_{t} = \left(
\begin{smallmatrix}
1 & 0 \\
0 & 0 \\
\end{smallmatrix}
\right)$,
and the inverse is done carried out using the Moore-Penrose pseudoinverse.

As done in Ref.~\cite{2013.PRA.Mower.DOQKD}, we shall assume that the excess-noise factors in the arrival-time and frequency measurements to be equal, i.e. $\xi_t = \xi_{\omega} = \xi$. With this assumption, we can make the simplification: $\eta_t = \eta_{\omega} = \eta$ and $\epsilon_t = \epsilon_{\omega} = \epsilon$. Thus, we can write Alice and Bob's covariance matrix after Eve's interaction as
\begin{equation}
\Gamma' =
\begin{pmatrix}
\gamma_{AA}			&	(1-\eta) \gamma_{AB}	\\
(1-\eta) \gamma_{BA}	&	(1+\epsilon) \gamma_{BB}
\end{pmatrix}.
\end{equation}
The relationship between the three noise parameters $\eta$, $\epsilon$, and $\xi$ is
\begin{equation}
\epsilon = \frac{-2\eta(d^2-1/4)+\xi}{d^2+1/4},
\end{equation}
where $d = \sigma_{\text{coh}}/\sigma_{\text{cor}}$ is the Schmidt number. After Alice and Bob estimate the value of $\xi$ from their data, they should then choose the values of $\eta$ and $\epsilon$ that maximize Eve's Holevo information. The range of possible $\eta$ and $\epsilon$ satisfy not only the relationship given above but also the following additional constraints:
\begin{inparaenum}[\itshape a\upshape)]
\item[(a)] Eve cannot increase Alice and Bob's mutual information by interacting with only Bob's photons due to the data processing inequality;
\item[(b)] the symplectic eigenvalues of the covariance matrix are greater than 1/2; and
\item[(c)] Eve can only degrade Alice and Bob's arrival-time correlations, i.e. $\Var\left[ T'_A-T'_B \right] \geqslant \Var\left[ T_A-T_B \right]$.
\end{inparaenum}

\section{Alice and Bob's mutual information}
\label{sec:shannon}
We assume that Alice and Bob only generate secure bits from their arrival-time measurements. During the reconciliation stage, Alice and Bob postselect frames in which each of them has at least one coincidence---either due to dark count or due to an actual photon. The probability for their postselecting a frame is given by~\eqref{eq:postselectProb}. In some of these postselected frames, either Alice or Bob may have registered more than one coincidence. To prevent Eve from exploiting multiple-coincidence frames, Alice and Bob replace such data with single coincidences chosen randomly from a Gaussian distribution whose variance equals the corresponding entry in the covariance matrix $\Gamma'$ plus the timing-jitter variance. Alice and Bob's arrival-time measurements therefore will derive from five different probability distributions~\cite{2014.PRL.Zhang.FransonQKD} as follows.
\begin{enumerate}
\item Bivariate Gaussian probability distribution with covariance matrix
	\begin{equation}
	\Lambda =
	\begin{pmatrix}
	\sigma_A^2		&	\Cov[T'_A,T'_B] \\
	\Cov[T'_A,T'_B]	&	\sigma_B^2
	\end{pmatrix},
	\end{equation}
	where $\Cov[T'_A,T'_B]$ means the covariance between $T'_A$ and $T'_B$, i.e. the top-left entry of the submatrix $\gamma'_{AB}$, $\sigma_A^2 = \Var[T'_A] + \sigma_J^2$, and $\sigma_B^2 = \Var[T'_B] + \sigma_J^2$. This case is a postselected frame in which Alice's source emitted one photon-pair and neither party had a dark count.

\item Independent Gaussian probability distributions with variances $\sigma_A^2$ and $\sigma_B^2$. This case is a postselected frame in which one of two situations occurred:
	\begin{inparaenum}[\itshape a\upshape)]
	\item[(a)] Alice's source emitted multiple photon-pairs, and Alice and Bob registered at least one coincidence; or
	\item[(b)] Alice's source emitted one photon-pair, and Alice and Bob registered a single coincidence with at least one of them also having a dark count. (There could be some correlations between Alice and Bob's measurements, but---being conservative---we are neglecting this possibility.)
	\end{inparaenum}

\item Alice's arrival time is a Gaussian random variable with variance $\sigma_A^2$, and Bob's arrival time is uniformly distributed over the measurement frame. This case is a postselected frame in which Alice detected at least one photon and Bob had a dark count without detecting photons.

\item Bob's arrival time is a Gaussian random variable with variance $\sigma_B^2$, and Alice's arrival time is uniformly distributed over the measurement frame. This case is a postselected frame in which Bob detected at least one photon and Alice had a dark count without detecting photons.

\item Both Alice and Bob's arrival times are uniformly distributed over the measurement frame. This is a postselected frame in which both Alice and Bob measured dark counts without detecting photons.
\end{enumerate}

The probability density functions for each of the above cases are
\begin{subequations}
\label{eq:probDist}
\begin{align}
p_{\left.T_A,T_B\right|1}(\left.t_A,t_B\right|1) &= p_{BG} (t_A,t_B;\Lambda), \\
p_{\left.T_A,T_B\right|2}(\left.t_A,t_B\right|2) &= p_G(t_A;\sigma_A^2) p_G(t_B;\sigma_B^2), \\
p_{\left.T_A,T_B\right|3}(\left.t_A,t_B\right|3) &= p_G(t_A;\sigma_A^2) p_U(t_B;T_f), \\
p_{\left.T_A,T_B\right|4}(\left.t_A,t_B\right|4) &= p_U(t_A;T_f) p_G(t_B;\sigma_B^2), \\
p_{\left.T_A,T_B\right|5}(\left.t_A,t_B\right|5) &= p_U(t_A;T_f) p_U(t_B;T_f),
\end{align}
\end{subequations}
where $p_{BG} (t_A,t_B;\Lambda)$ is a bivariate Gaussian probability density function with zero means and covariance matrix $\Lambda$; $p_G(t;\sigma^2)$ is a Gaussian probability density function with zero mean and variance $\sigma^2$; and $p_U(t;T_f)$ is a uniform probability density function over the interval $[-T_f/2,T_f/2]$.

Moreover, the probabilities for each of the cases discussed above, given that a particular frame has been postselected, are
{\allowdisplaybreaks
\begin{subequations}
\begin{align}
\pi_1 &= \mu e^{-\mu} \eta_A \eta_B \eta_P (1-p_d)^2 / P_{\mu}, \\
\pi_2 &= \sum^{\infty}_{n=2} \frac{\mu^n e^{-\mu} }{n! P_{\mu}} \left[ 1 - (1-\eta_A)^n \right] \left[ 1 - (1-\eta_B \eta_P)^n \right] \nonumber \\
	&\qquad + \frac{\mu e^{-\mu}}{P_{\mu}} \eta_A \eta_B \eta_P p_d (2-p_d), \\
\pi_3 &= \sum^{\infty}_{n=1} \frac{\mu^n e^{-\mu}}{n! P_{\mu}} \left[ 1 - (1-\eta_A)^n \right] \left[ p_d (1-\eta_B \eta_P)^n \right], \\
\pi_4 &= \sum^{\infty}_{n=1} \frac{\mu^n e^{-\mu}}{n! P_{\mu}} \left[ p_d (1-\eta_A)^n \right] \left[ 1- (1-\eta_B \eta_P)^n \right], \\
\pi_5 &= \sum^{\infty}_{n=0} \frac{\mu^n e^{-\mu}}{n! P_{\mu}} p_d^2 (1-\eta_A)^n (1-\eta_B \eta_P)^n,
\end{align}
\end{subequations}}
where $\eta_A$, $\eta_B$, $\eta_P$ and $p_d$ have been defined in Sec.~\ref{sec:postselectionProbability}.

The conditional probability density functions defined above, as well as their occurrence probabilities, allow us to define the arrival-time joint probability density function:
\begin{equation}
p_{T_A,T_B}(t_A,t_B) = \sum^{5}_{i=1} \pi_i \; p_{\left.T_A,T_B\right|i}(\left.t_A,t_B\right|i).
\end{equation}
Using this joint probability density function, we can calculate Alice and Bob's mutual information via
\begin{equation}
\label{eq:ShannonInfoAB}
\begin{aligned}
I(A;B)_{\mu} &=\int \intd{t_A} \intd{t_B} \; p_{T_A,T_B}(t_A,t_B) \\
	&\qquad \times \log_2 \left( \frac{p_{T_A,T_B}(t_A,t_B)}{p_{T_A}(t_A) p_{T_B}(t_B)} \right),
\end{aligned}
\end{equation}
where $p_{T_A}(t_A) = \int \intd{t_B} \: p_{T_A,T_B}(t_A,t_B)$ and $p_{T_B}(t_B) = \int \intd{t_A} \: p_{T_A,T_B}(t_A,t_B)$ are the marginal probability density functions.

It is important to note that when the detector timing jitter $\sigma_J$ exceeds the correlation time $\sigma_{\text{cor}}$, Alice and Bob's mutual information $I(A;B)$ cannot approach its limit of $\log_2{d}$. In this case, the WDM~\cite{2011.ArXiv.Mower.DWDM-QKD} that makes $\sigma_{\text{cor}}$ in each WDM channel comparable to $\sigma_J$ should be applied, and secure-key must be obtained from both arrival-time and frequency measurements.


\bibliography{References/references}
\end{document}